
\input harvmac.tex

\let\bar=\overline

\def\suii{$\widehat{su(2)_q}$}
\let\<=\langle
\let\>=\rangle
\let\e=\epsilon
\let\ep=\epsilon
\let\tht=\theta
\let\t=\theta

\let\p=\prime

\noblackbox
\pretolerance=750
\lref\fs{P. Fendley and H. Saleur, Nucl. Phys. B388 (1992) 609.}
\lref\YY{C.N. Yang and C.P. Yang, Phys. Rev. 150 (1966) 321, 327;
151 (1966) 258.}
\lref\fmvw{P. Fendley, S.D. Mathur, C. Vafa and N.P. Warner,
Phys. Lett. B243 (1990) 257.}
\lref\pk{P. Fendley and K. Intriligator, Nucl. Phys. B372 (1992) 533.}
\lref\pkii{P. Fendley and K. Intriligator, Nucl. Phys. B380 (1992) 265.}
\lref\polya{A.M. Polyakov, {\it Gauge Fields and Strings} (Harwood,1987).}
\lref\witti{E. Witten, Nucl. Phys. B202 (1982) 253.}
\lref\flmw{P. Fendley, W. Lerche, S.D. Mathur and N.P. Warner,
Nucl. Phys. B348 (1991) 66; P. Mathieu and M. Walton, Phys. Lett. 254B (1991)
106.}
\lref\rZandZ{A.B. Zamolodchikov and Al.B. Zamolodchikov, Ann.
Phys. 120 (1980) 253.}
\lref\rAZi{A.B. Zamolodchikov, Adv. Stud. Pure Math. 19 (1989) 1.}
\lref\tbaref{Al.B. Zamolodchikov, Nucl. Phys. B342 (1990) 695.}
\lref\BR{V.V.  Bazhanov
and N.Yu.  Reshetikhin, Int. J.  Mod. Phys. A4 (1989) 115.}
\lref\dhoker{E. D'Hoker and D.H. Phong, Rev. Mod. Phys. 60 (1988) 917.}
\lref\RSOS{A.B. Zamolodchikov, Landau Institute preprint,
September 1989.}
\lref\ku{K. Kobayashi and T. Uematsu, Phys. Lett. B275 (1992) 361.}
\lref\lnw{A. LeClair, D. Nemeschansky and N. Warner, Nucl. Phys. B390 (1993)
653. }
\lref\BCN{H.W.J. Bl\"ote, J.L. Cardy and M.P. Nightingale, Phys. Rev.
Lett. 56 (1986) 742; I. Affleck, Phys. Rev. Lett. 56 (1986) 746.}
\lref\jnw{G. Japaridze, A. Nersesyan and P. Wiegmann,
Nucl. Phys. B230 [FS10] (1984)
511; P. Wiegmann, Phys. Lett. B152 (1985) 209.}
\lref\hasen{
P. Hasenfratz, M. Maggiore and F. Niedermayer, Phys. Lett. B245 (1990) 522;
P. Hasenfratz and F. Niedermayer, Phys. Lett. B245 (1990) 529.}
\lref\sausage{V.A. Fateev, E. Onofri and Al. B. Zamolodchikov, ``The
sausage model (integrable deformations of O(3) sigma model)'', LPTHE
preprint 92-46.}
\lref\smirff{F. Smirnov, {\it Form-factors in Completely Integrable Models of
Quantum Field Theory},  Singapore: World Scientific (1992).}
\lref\BL{D. Bernard and A. LeClair, Phys. Lett. 247B (1990) 309.}
\lref\cfiv{S. Cecotti, P. Fendley, K. Intriligator, and C. Vafa, Nucl.
Phys. B386 (1992) 405.}
\lref\FSZii{P. Fendley, H. Saleur and Al.B. Zamolodchikov,
``Massless Flows II: the exact $S$-matrix approach'', hepth@xxx/9304051, to
appear in Int. J. Mod. Phys. A.}
\lref\afl{N. Andrei, K. Furuya, and J. Lowenstein, Rev. Mod. Phys. 55
(1983) 331.}

\Title{\vbox{\baselineskip12pt\hbox{RU-93-26}
\hbox{BUHEP-93-16}}}
{\vbox{\centerline{Central charges without finite-size effects}}}
\vglue .5cm
\centerline{Paul Fendley$^1$\footnote{$^\dagger$}
{address after Sep.\ 1, 1993: Department of Physics,
University of Southern California, Los Angeles, CA 90089}
 and Ken
Intriligator$^2$}
\vglue .5cm
\vglue .3cm
\centerline{$^1 $Department of Physics, Boston University}
\centerline{590 Commonwealth Avenue, Boston, MA 02215, USA}
\centerline{\it fendley@ryan.bu.edu}
\vglue .3cm
\centerline{$^2$ Department of Physics}
\centerline{Rutgers University, Piscataway, NJ 08855, USA}
\centerline{\it keni@physics.rutgers.edu}
\vglue 1cm
We show how to obtain the ultraviolet central charge from the exact $S$-matrix
for a wide variety of models with a $U(1)$ symmetry. This is done by coupling
the $U(1)$ current $J$ to a background field. In an $N$=2 superconformal
theory with $J$  the fermion number current, the OPE of $J$ with itself
and hence the free energy are proportional to $c$. By deforming the
supersymmetry into affine \suii quantum-group symmetry, this result can be
generalized to many $U(1)$-invariant theories, including the $N$=0 and $N$=1
sine-Gordon models and the $SU(2)_k$ WZW models.  This provides a consistency
check on a conjectured $S$-matrix completely independent of the finite-size
effects expressed in terms of dilogarithms resulting
from the thermodynamic Bethe ansatz.

\Date{7/93}
One way of solving integrable two-dimensional field theories is to put the
model on the lattice and diagonalize the quantum Hamiltonian by using the
Bethe ansatz, much in the same way that one solves a quantum spin chain.
However, another approach has become popular in recent years: one uses the
enormous constraints of an integrable theory to conjecture the exact
$S$-matrix \rZandZ. Instead of dealing with field theories, one studies
relativistic quantum particles. {}From the exact $S$-matrix, one can derive
exact form factors \smirff\ and extract correlators. One can also derive the
free energy at non-zero temperature from the thermodynamic Bethe ansatz (TBA)
\tbaref.  An advantage of the exact $S$-matrix approach (and the main reason so
much effort has been devoted to it in recent years) is that models can be
solved by defining them as conformal field theories perturbed away from the
critical point, without knowing the action or Hamiltonian explicitly \rAZi.

When conjecturing an exact $S$-matrix, it is important to check that it has
the right properties by comparing results with conventional Lagrangian
perturbation theory, conformal perturbation theory or numerical simulations.
One very unambiguous check is to calculate the Casimir energy from the TBA (by
changing the roles of space and time, finite-temperature effects become
finite-size effects), because in a critical theory, it is proportional to the
central charge $c$ \BCN.  Many such calculations have been done by now (too
many to refer to them all).  The result is given in terms of sums of
dilogarithms, and to obtain the final result, one must use or derive some
marvelous dilogarithm identities.

The purpose of this letter is to show that there is another simple quantity
which is proportional to the central charge in any theory with $N$=2
supersymmetry, and many theories with a $U(1)$ symmetry.  We show how to
calculate this quantity from the $S$-matrix, giving a check on the $S$-matrix.
This calculation is done at zero temperature but requires adding a background
field coupled to a conserved charge.

In the past, the effects of a background field coupled to a $U(1)$ current
have been used to relate parameters in the Lagrangian to the $S$-matrix
\refs{\hasen,\sausage,\FSZii}. Here we will see how there is an even bigger
payoff in theories with $N$=2 supersymmetry, where the $U(1)$ is the fermion
number. Because the chiral anomaly in an $N$=2 theory is proportional to the
central charge $c$, $c$ is determined directly from this calculation,
independently of any details of the specific Lagrangian.  For example, no
renormalization effects need be studied.  Moreover, this is not the only case
where this observation is useful, because one can deform the $N$=2
supersymmetry into an affine
\suii\ quantum-group symmetry while keeping the model integrable \BL.
(If one bosonizes the $U(1)$ current, this
is equivalent to changing the radius of the boson.)  Many integrable models
with a $U(1)$ symmetry can be obtained in this manner, so this provides a
strong check on these $S$-matrices as well.

Consider a theory with a $U(1)$ symmetry generated by a current $J_{\mu}$.  We
write the conservation law as $\del J_L+\bar \del J_R=0$.  The corresponding
conserved charge $Q$ can be coupled to a background ``electric'' field $A$
which we take to be constant to ensure that we don't destroy the integrability
of our theory.  This modifies the Hamiltonian to be $H=H_0+QA$.  Because $A$
has dimensions of mass, the strength of $A$ controls the position of our
theory on its renormalization group trajectory and, in particular, in the
limit of large $A$ the theory is driven to the UV.  The ground-state energy of
a theory in the limit of large background field $A$ thus is related to a
property of its UV conformal fixed point theory.

We first show that the ground-state energy in a background field is related to
the chiral anomaly of the $U(1)$ current.  Suppose at a conformal fixed point
of the flow that the left and right charges are separately conserved, i.e.
$\del J_L$=$\bar\del J_R$=0.  We can then couple independent background fields
to the left and right charges, modifying the hamiltonian as
$H=H_0+A_LQ_L+A_RQ_R$.  Off the fixed point we will need to set $A_L=A_R$
because only $Q=Q_L+Q_R$ is conserved, but for the moment we can treat them
independently.  The action is modified by the background fields as
\eqn\action{S= S_0 +\int {d^2z\over 2\pi} (A_LJ_L+A_RJ_R).}
We regulate the theory by taking spacetime to be a torus of euclidean time
$\beta$ and spatial length $L$; at the end we take the volume to infinity.  In
the large $\beta$ limit we have $Tre^{-\beta H}\rightarrow e^{-\beta E(A)}$,
where $E(A)$ is the ground-state energy in background field $A$.  Evaluating
the partition function as a path integral using \action , we find in this
limit
\eqn\fey{e^{-\beta (E(A)-E(0))}=\< e^{-\int {d^2z\over 2\pi}
 (A_LJ_L+A_RJ_R)}\>,}
where the correlator is evaluated in the theory without the background fields,
i.e. using $S_0$.  By Wick's theorem, \fey\ yields
\eqn\wt{ E(A)-E(0)={1\over 2\beta} \int {d^2z\over 2\pi} \int
{d^2w\over 2\pi}
\<(A_LJ_L(z)+A_RJ_R(z))(A_LJ_L(w)+A_RJ_R(w))\>.}

The operator products of the currents appearing in \wt\
are generally of the form
\eqn\opes{\eqalign{\<J_L(z)J_L(w)\>={\kappa \over (z-w)^2}+
{2\pi\kappa \over V},
\qquad & \<J_R(z)J_R(w)\>={\kappa\over (\bar z-\bar w)^2}+
{2\pi\kappa\over V},\cr
\<J_L(z)J_R(w)\>=2\pi &\kappa \delta ^2 (z-w) -{2\pi\kappa\over V},}}
where $\kappa$ is a constant measuring the strength of the chiral anomaly and
$V=\beta L$ is the area of the worldsheet.  These OPEs require some
explanation.  In the infinite volume limit we can use the planar expressions
for the OPEs, which are the first terms in the above equations.  The contact
term in $\<J_LJ_R\>$ is required for vector current conservation $\del
J_L+\bar\del J_R=0$, for example in $\<(\del J_L+\bar\del J_R)J_L\>$.  The
constant $1/V$ terms in \opes\ reflect the need to absorb the zero mode
associated with the normalizable constant modes existing in finite volume.
The above expressions can be obtained using bosonization by taking derivatives
of the propagator of a free boson on the torus.  {}From the explicit form of
this propagator (for example, see \dhoker ) we find two terms, one involving
the theta function $\t _1$ which results in the first terms in the above
expressions and another (necessary for modular invariance) yielding the
$1/V$ terms.

In our infinite volume limit one might be tempted to drop the $1/V$ terms in
\opes .  Plugging \opes\ into \wt\ we see that we can do so if
the background fields go to zero at infinity sufficiently rapidly so that
$V^{-1}\int d^2z A(z)\rightarrow 0$.  The remaining pieces yield the familiar
\polya\ result $\kappa \int (d^2z/2\pi) F(\del\bar\del )^{-1}F$,
where $F=\del A_L-\bar\del A_R$ is the gauge-invariant field strength.  In
this expression left and right movers are coupled through a $A_LA_R$ term
coming from the $\delta ^2(z-w)$ in \opes. This is the familar statement of
the chiral anomaly: by imposing vector current conservation we ensure
gauge invariance at the expense of left-right decoupling.  This is why
$\kappa$ measures the chiral anomaly.

We, however, are interested in constant background fields so we clearly
need to keep the $1/V$ terms in \opes .  The $1/(z-w)^2$ and $1/(\bar z
-\bar w)^2$ pieces vanish in the integrals and we are left with the final
result for the energy density due to the constant background fields
\eqn\confenergy{{\cal E}(A) \equiv {E(A)-E(0)\over L}=-{\kappa \over
4\pi}(A_L^2+A_R^2).}
The fact that the $A_LA_R$ term dropped out is sensible: the chiral anomaly is
proportional to the derivatives of the background fields so for truly constant
fields it vanishes and the left and right movers should decouple.  To make
contact with our theory away from the critical point we set $A_L$=$A_R$=$A$
and \confenergy\ becomes
\eqn\gsedk{{\cal E}(A\rightarrow \infty)=-{\kappa _{uv} \over 2\pi}A^2,}
where $\kappa _{uv}$ is the chiral anomaly of the UV critical theory as
defined in \opes.  For large but finite $A$, \gsedk\ will have subleading
terms in $(M/A)$, where $M$ is the mass scale.

In order to make the connection between $\kappa$ and the central charge of the
UV fixed point we note that in many integrable models the $U(1)$ symmetry is
part of a \suii\ symmetry ($N$=2 supersymmetry is a special case of the
quantum-group symmetry at $q^2=-1$.). Our convention will be to always
normalize the $U(1)$ charge so that the $Q^{\pm}$ generators of \suii\ have
charge $\pm 1$.  We can bosonize the $U(1)$ symmetry as:
\eqn\boson{J_L={\lambda\over 2\pi R}
 \partial \phi, \qquad J_R=-{\lambda\over 2\pi
R}\bar \partial \phi,}
where $\phi \sim\phi +2\pi R$ and the normalization constant $\lambda$ is
independent of $R$ and determined by our above convention.  From this we find
$\kappa=(\lambda/2\pi R)^2$.  In an $N$=2 theory, the supersymmetry algebra
requires that $\kappa=c/3$.  Now the crucial point is that we can change the
theory by changing the value of the boson's radius $R$; this changes $q$.
Because changing the radius $R$ doesn't change the central charge, this yields
a general relation between $\kappa$ and $c$, namely
\eqn\gsed{{\cal E}(A\rightarrow \infty)=-{c_{uv}\over 6\pi}
{R^2_{N=2}\over R^2}A^2.}

We now show how to calculate the ground-state energy \gsed\ directly from the
$S$-matrix, giving the promised check. Turning on the background field has the
effect of changing the ground state. If one were working in field theory or in
a lattice model, this would change the Dirac or Fermi sea. In our exact
particle description, a similar thing happens: the ground state no longer is
the empty state --- it has a sea of real particles.  We use rapidity variables
so that a particle of charge $q$ and mass $m$ has energy $qA + m\cosh\t$.
When $qA$ is negative, for small enough $\t$ this particle has negative energy
and it will lower the energy as part of the ground state. In all known
integrable models save that of a free boson, the particles behave as if they
were fermions; i.e.\ there is only one particle to a level.  Thus in the
ground state such particles fill all levels with $|\t|$ less than some
number $B$ (if the particles did not interact, we would have $|q A|=m \cosh
B$).  Any type of particle with $q_aA<0$ can appear in the ground state, so we
label them by the index $a$. Defining the associated ground-state densities by
$\rho_a(\t)$, the energy density of the ground state in a box of length $L$ is
thus
\eqn\gsi{{\cal E}(A)= {1\over L}\sum_a
 \int_{-B_a}^{B_a} d\t\ \rho_a(\t)(-|q_a A|+m_a \cosh\t). }

In a general interacting theory, one must resort to perturbation theory.
However, the exact $S$-matrix allows us to derive equations for the
$\rho_a(\t)$ and $B_a$. These follow simply from putting the system in a box
of length $L$ and imposing periodic boundary conditions; similar calculations
have been done in lattice models \refs{\YY,\afl,\jnw}. It is possible to study
this constraint because the scattering is completely elastic; the individual
momenta remain the same in a collision. The scattering need not be diagonal in
general --- internal labels can change. When the scattering is diagonal among
particles in the ground state (i.e.\ the only scattering is of the type $a(\t
_1)b(\t _2)\rightarrow S_{ab}(\t _1-\t _2)b(\t _2)a(\t _1)$), the momentum of
the a particle of type $a_k$ is quantized via
\eqn\quant{e^{im_k\sinh \t _kL}\prod_{j\ne k} S_{a_ka_j}(\theta_k-\theta_j) =
1,}
an interacting-model generalization of the one-particle relation $p_k=2\pi
n_k/L$.  In the large $L$ limit, taking the derivative of the log of
\quant\ gives the distribution $P_a(\t )$ of
allowed rapidity levels for particles of species $a$. Since we are at zero
temperature, the particles fill all allowed levels up to $B_a$, i.e.
$\rho_a(\t)=P_a(\t)$ for $|\t|< B_a$, giving
\eqn\forrho{\rho_a(\tht)= {mL\over 2\pi} \cosh\theta+
\sum_b \int_{-B_a}^{B_a} d\tht'\rho_b(\tht')\phi_{ab}(\theta-\theta'),}
where $$\phi_{ab}(\tht)\equiv -{i\over 2\pi}
{\del\ln{S_{ab}(\tht)}\over\del\tht}.$$
The equations \forrho\ and \gsi\ and the equations for $B_a$ (given by
minimizing the energy with respect to $B_a$) can be put in a more convenient
form by defining the ``dressed'' particle energies $\ep_a(\t)$ as
\eqn\fore{\e _a(\t )=|q_aA| - m_a\cosh \t  +\sum _b
\int_{-B_b}^{B_b} d\t ^{\p}\phi_{ab}(\t -\t ^{\p})\e _b(\t ^{\p}).}
Substituting this into \gsi\ and using \forrho\ one finds that
\eqn\gsii{{\cal E}(A)= -\sum_a{m_a\over 2\pi}\int_{-B_a}^{B_a}d\t\ \cosh\t
\ \e_a(\t).}
It is obvious that the $B_a$ are determined by the boundary conditions
$\ep_a(\pm B_a)=0$.
As we will discuss below, the same equations are valid even when the
scattering among particles in the ground state is not diagonal if
the appropriate zero-mass, zero-charge ``pseudoparticles'' are included.

The relations \fore --\gsii\
are not solvable in general, but the results can be written as a series in
$m_a/A$ by using a generalized Weiner-Hopf technique \jnw.  However, in the
conformal limit where we make contact with the field dependence derived in
\gsed, they can be solved. Following \sausage, we define $\bar \ep_a$ by
\eqn\foree{\bar\e _a(\t )=|q_aA| - {m_a\over 2} e^\t  +\sum _b
\int_{-\infty}^{B_b}
d\t ^{\p}\phi_{ab}(\t -\t ^{\p})\bar\e _b(\t ^{\p}).}
To reach the ultraviolet limit, we take $A$ very large. In this case, the
$B_a$ are very large and the energy \gsii\  can be written as
\eqn\gsiii{{\cal E}(A)\approx -\sum_a{m_a\over 2\pi}\int_{-\infty}^{B_a}
d\t\ e^{\t} \bar\e_a(\t).}
Taking the derivative of \foree\ with respect to $\t$ and using it to get rid
of the $m_a\exp\t$ in
\gsiii, after using \foree\ again one finds
\eqn\gsiv{{\cal E}(A\rightarrow \infty )
= -\sum_a {|q_aA|\over 2 \pi} \bar\ep_a(-\infty)}
It follows from \foree\ that the $\bar\ep_a(-\infty)$ follow from the
matrix equation
\eqn\epinf{(\delta_{ab}-N_{ab})\bar\ep_b(-\infty)=|q_a A|.}
where we define $N_{ab}\equiv \int_{-\infty}^{\infty} d\t \phi_{ab}(\t)$.  We
thus have arrived at a very simple answer for the ground state energy at the
ultraviolet fixed point.  It is given very simply in term of the charges and
the $N_{ab}$, and the latter follow simply from the $S$-matrix. The
ground-state energy \gsiv\ must of course must agree with
\gsed, giving a simple check on the $S$-matrix.

We devote the remainder of this letter to providing examples. First we discuss
the sine-Gordon model,
\eqn\sg{S= \int d^2 z \left[ \half (\del \phi)^2 + \mu^2 \cos\beta\phi\right],}
repeating the analysis of \FSZii.  The conserved charge is proportional to the
topological charge and the corresponding currents may be written as \boson\
where the radius is that of the SG boson $R=\beta ^{-1}$.  For $\beta^2\ge
4\pi$ the entire spectrum consists of a soliton and antisoliton.  There
is a \suii\ quantum group symmetry with $q=-\exp (-i\pi/t)$
where the parameter $t$, which appears in the S-matrix element written
below, is related to $\beta$ by
\eqn\betasq{\beta^2= 8\pi {t\over t+1}.}
In our normalization where the quantum-group generators $Q^{\pm}$ have charge
$\pm 1$, the soliton has charge $\half$, while the antisoliton has charge
$-\half$ \BL.  The $N$=2 value of $\beta$ ($t$=2) is $\beta^2=(2/3)8\pi$;
there the charge is fermion number and the fact that the solitons have fermion
number $\pm \half$ \pk\ is the phenomenon of fractional fermion number.  When
$A>0$, the ground state will fill with antisolitons. These scatter among
themselves with the $S$-matrix element \rZandZ
\eqn\saa{\phi(\theta)={1\over 2\pi i}{d\over d\theta}\log S(\theta)=
\int{e^{i\omega\theta}\sinh{\pi(t-1)\omega\over 2}\over
2\cosh{\pi\omega\over 2}\sinh{\pi t\omega\over 2}}{d\omega\over 2\pi},}

There is thus only one element in $N_{ab}$, which is $(t-1)/2t$. Using \epinf,
we see that $\bar\ep(-\infty)=tA/(t+1)$, and plugging this into \gsiv\ gives
the ground-state energy density at the critical point
$${\cal E}(A\rightarrow \infty )
= -{A^2\over 4\pi}{t\over (t+1)}= -{1\over 6\pi}
{\beta^2\over \beta_{N=2}^2}A^2$$
giving the correct result $c=1$ and the correct $\beta$ dependence of \gsed .
This is thus a very non-trivial check on the sine-Gordon $S$-matrix and the
relation \betasq.  In fact, this calculation can be alternatively viewed as a
way of deriving the relation \betasq\ between parameters $t$ in the $S$-matrix
and $\beta$ in the Lagrangian.


Our next examples are the ``fractional supersymmetric sine-Gordon'' models \BL
.  The soliton content of these theories is described by $k+1$ vacua on a line
with ``RSOS'' solitons interpolating between adjacent vacua.  For each pair of
vacua $(j,j\pm 1)$ there is a doublet of solitons with charge $\pm \half$.
The conjectured $S$-matrix is a tensor product of the $k-th$ RSOS \RSOS\
$S$-matrix governing scattering in the soliton labels and the ordinary
sine-Gordon $S$-matrix governing scattering in the doublet labels \refs{\BL,
\pk}.  When $A>0$, all the negatively-charged particles can be in the sea.
While scattering among these particles is diagonal in the ordinary sine-Gordon
labels, it is not diagonal in the soliton labels. For example, the
two-particle state with ($j,j+1$) and ($j+1,j$) solitons can scatter into
itself or into ($j,j-1$) and ($j-1,j$).  To quantize the momenta, one must
find the analog of \quant\ in this non-diagonal case; in other words, one must
find the eigenvalues for bringing one particle through an ensemble of the
others. This is precisely what one does in a traditional Bethe ansatz
calculation. In this case, the calculation has already been done in the
minimal-model guise \refs{\BR,\fs}. As in the TBA, the effect of the soliton
degrees of freedom is to introduce additional ``pseudoparticles''.  They enter
into the analysis just like the real particles, but have zero mass and zero
charge.  The result is that there are $k-1$ pseudoparticles (labelled $1\dots
k-1$) and one ``real'' particle with charge $-\half$ (labelled by $k$).  This
calculation gives the kernel $\phi_{a,a\pm 1}(\t)=1/(2\pi\cosh\t)$ with all
others zero; the only modification for our full tensor-product case is that
there is an additional piece coming from the sine-Gordon $S$-matrix.  Since
this part of the scattering is diagonal, $\phi_{kk}$ is given by
\saa . This yields $N_{a,a\pm 1}= \half$, $N_{kk}=(t-1)/2t$, and the
rest are zero. The solution of \epinf\ is
$$\bar\ep_a(-\infty)= {atA\over (k+t)}.$$
Only the real particle $k$ has charge, so it is the only one which
contributes to the energy \gsiv
\eqn\fssg{{\cal E}(A\rightarrow \infty)=-{kt\over 2(k+t)}{A^2\over 2\pi}.}

When $t$=2 these theories are $N$=2 supersymmetric and thought to describe the
integrable theories obtained by perturbing the $k$-th $N$=2 minimal model in
its least relevant operator \pk .  Setting $t$=2 in \fssg\ gives the correct
central charge $3k/(k+2)$.  Comparing \gsed\ with
\fssg , we also find that the radius $R$ of the bosonized $U(1)$ current
is related to the quantum-group parameter $t$ by
$${R^2_{N=2}\over R^2}={t(k+2)\over 2(t+k)}.$$
Setting $R_{N=2}^{-2}=8\pi (2/k+2)$, this is in agreement with the relation
derived using the quantum-group symmetry in \BL\ (eqn.\ 3.12 there).  The
$k$=2 case corresponds to the $N$=1 sine-Gordon model.  Another special case
is the $t\rightarrow \infty$ limit, where the quantum-group symmetry becomes
ordinary affine $\widehat{SU(2)}_k$; this corresponds to the $SU(2)_k$ WZW
model perturbed by the current-current interaction.

Another integrable perturbation of the $N$=2 discrete series is by the most
relevant operator \refs{\fmvw,\flmw}.  The vacua can be drawn as the $k+1$
vertices of a regular polygon, labelled $1\dots k+1$ counterclockwise, and the
spectrum consists of solitons interpolating between any pair of vacua.  Each
such soliton is a doublet; the charges of the doublet interpolating between
the vacua $j$ and $(j+r)$ mod $k+1$ are given by $r/(k+1)$ and $r/(k+1)-1$
\pkii.  With a positive magnetic field, one would expect that any
negatively-charged particle (the member of each doublet with charge $r/(k+1)
-1$, which we can label by $r=1\dots k$) would be allowed in the sea.  These
scatter diagonally among themselves (so we don't need pseudoparticles) with
the $S$-matrix element $\tilde a_{r,s}(\t )$ obtained in \pkii .  We need only
the $\tilde a_{r,1}$, which give
$$N_{r,1}=\delta _{r,1}-{(k+1-r)(k+2)\over (k+1)^2},$$
%
%
%
%
from which it follows that the solution of \epinf\ is $\bar\ep_r(-\infty)=0$
if $r\ne 1$ and $\bar\ep_1(-\infty)=(k+1)A/(k+2)$. Using these in \gsiv\ gives
\eqn\gsmin{{\cal E}(A\rightarrow \infty )= -{k\over k+2} {A^2\over 2\pi}.}
Only the $r$=1 particle with the most-negative charge $-k/(k+1)$ contributes
to this leading term in the energy; in fact it can be shown that the other
particles do not appear in the sea at all.

Comparing \gsmin\ with \gsed, we again obtain the correct central charge of
$3k/(k+2)$. This provides a check beyond the TBA \pkii\ that the solitons form
the entire spectrum of the theory, and the $S$-matrix given there is the
correct one.  As for deforming the $N$=2 symmetry into a quantum-group
symmetry, this would presumably give an affine Toda field theory with
imaginary coupling constant and a background charge \refs{\flmw,\lnw}, but
these theories have not been studied a great deal. However, this calculation
should provide the relation of the Toda coupling constant to the $S$-matrix
parameter.

We also note that result \gsed\ can be continued to imaginary background
field.  Setting $A=i\alpha T$ and taking the temperature $T\rightarrow
\infty$, we reach the UV critical point. Multiplying the energy density from
\gsed\ by the box length $1/T$ (associated with finite temperature) to obtain
an energy and adding the $A$=0 energy which follows from \BCN , we obtain the
total ground state energy at the critical point
$$E(T)=-{\pi c T\over 6}(1- {R^2_{N=2}\over R^2} {\alpha^2\over \pi^2}).$$
This result is in perfect agreement with several specific examples \refs{ \pk,
\fs}\ obtained by including imaginary chemical potentials
in the TBA calculation and using dilogarithm identities.  For a $N$=2 system,
where $(R_{N=2}/R)$=1, this is a familiar result for fermion boundary
conditions twisted by $e^{i\alpha F}$.

This of course does not exhaust the set of conjectured $N$=2 $S$-matrices.  We
have checked that the conjectured $S$-matrices for the $N$=2 sine-Gordon model
\ku\ also give the correct answer.  There are models with more general soliton
structure discussed in \refs{\pkii,\lnw} which have not been checked here or
via the TBA. The background-field calculation described here will almost
certainly turn out to be easier, because one only needs to diagonalize the
$S$-matrix for negatively-charged particles, not the full spectrum. In
addition, there is no need for dilogarithm identities (although some may feel
this is a disadvantage!).  Another advantage of this method is that being at
zero temperature, we do not need to worry about finite-size effects. This is
useful because there is at least one case where the presence of level
crossings in finite volume means that calculating anything from the
finite-size effects of the TBA is suspect \FSZii.

\centerline{\bf Acknowledgements}

We would like to thank Mike Douglas for a helpful discussion.
P.F.\ was supported by DOE grant DEAC02-89ER-40509, K.I.  by
DE-FG05-90ER40559.

\listrefs

\end